\documentclass[12pt]{article}
\usepackage{latexsym,amsmath,amssymb,amsbsy,graphicx}
\usepackage[utf8]{inputenc}
\usepackage[T2A]{fontenc}
\usepackage[english]{babel}
\usepackage[space]{cite} 

\makeatletter\def\@biblabel#1{\hfill#1.}\makeatother

\allowdisplaybreaks
\usepackage{verbatim}  
\usepackage{color}     
\usepackage{subfigure} 
\usepackage{hyperref}  
\usepackage{graphics}
\usepackage{bm}
\usepackage{ulem}
\usepackage{lineno}
\usepackage{booktabs}
\usepackage{dcolumn}

\begin{document}

	\noindent\begin{minipage}{\textwidth}
		\begin{center}
			
			
			{\Large{The nucleon resonance structure from exclusive $\pi^+\pi^-p$ photo-/electroproduction off protons}}\\[9pt]
			
			{\large V.D.~Burkert$^{1}$, V.I.~Mokeev$^{1}$, B.S.~Ishkhanov$^{2}$}\\[6pt]
			
			
			\parbox{.96\textwidth}{\centering\small\it
				$^1$Thomas Jefferson National Accelerator Facility, Newport News, Virginia 23606, USA\\
				$^2$Skobeltsyn Nuclear Physics Institute and Physics Department at Moscow State University, 
				119899 Moscow, Russia\\
				\ E-mail: bsi@depni.sinp.msu.ru\\[1cc] 
			}
			\parbox{.96\textwidth}{\centering\small Статья поступила --.--.2019, подписана в печать --.--.2019.} 

		\end{center}
		
		{\parindent5mm 
			
			The results on photo- and electroexcitation amplitudes of most nucleon resonances in the mass range up to 2.0 GeV determined from the CLAS experimental data on exclusive $\pi^+\pi^-p$ photo-/electroproduction off protons in collaboration between the Jefferson Lab and Moscow State University are presented. The first and only available results on electroexcitation amplitudes from CLAS in a wide range of photon virtualities $Q^2$ $<$ 5.0 GeV$^2$ revealed the nucleon resonance structure as a complex interplay between the inner core of three dressed quarks and external meson-baryon cloud. These results shed light on the strong QCD dynamics which underlines the generation of excited nucleon states of different structural features from confined quarks and gluons. The future prospects of these studies in the new era of experiments with the CLAS12 detector, which started successfully in Spring of 2018, are outlined.

			\vspace{2pt}\par}


		\vspace{1pt}\par
%
		\vspace{1pt}\par
	\end{minipage}

\section*{Introduction}
Studies of the excited 
nucleon ($N^*$) structure offer unique opportunities to explore many facets of strong QCD dynamics as it generates 
various excited nucleons with different quantum numbers of distinctively different structure~\cite{Bu12,BurkertRoberts,Roberts18}. Theoretical analyses of the nucleon resonance electroexcitation amplitudes ($\gamma_vpN^*$ electrocouplings) within the framework 
of different quark models have demonstrated that almost all quark models are capable of reasonably well describing the nucleon elastic form factors by adjusting their parameters, but they predict a  distinctly different evolution of the $\gamma_vpN^*$ electrocouplings with photon virtualities $Q^2$ for excited nucleon states~\cite{Bu12,Az12,Az17,Br18,Ra18,Ob11,Ob16,Lyub17,San15}. Advances in the continuum-QCD Dyson-Schwinger equation (DSE) approach make it possible for the first time to explore the strong QCD dynamics behind the generation of the dominant part of hadron mass~\cite{Seg14,Seg15} using the results on the nucleon elastic form factors and the $Q^2$-evolution of different resonance electrocouplings. Mapping out the dressed quark mass function addresses the most challenging open problems of hadron physics on the nature of hadron mass and quark-gluon confinement. It makes the exploration of the ground and excited nucleon structure using the experimental data on nucleon elastic form factors and resonance electrocouplings particularly important for the understanding of the strong QCD dynamics behind hadron generation.

In this paper we present the results on the studies of $\gamma_vpN^*$ electrocouplings from the CLAS detector data on exclusive charged double pion photo- and electroproduction off protons. In the last decade, the exploration of the resonance structure from the data of this exclusive channel represents the major direction in collaborative efforts between Hall B at Jefferson Lab and Skobeltsyn Nuclear Physics Institute, Physics Department at Moscow State University \cite{Mok18}. We also outline the future prospects of these studies in the new era of experiments with the CLAS12 detector \cite{Bu19}. These experiments already started successfully in Spring of 2018.

\section{The current status}
\subsection{CLAS experimental data on $\pi^+\pi^-p$ electroproduction off protons and the Jefferson Lab-Moscow State University model for the resonance parameter extraction} 

Exclusive $\pi^+\pi^-p$ photo- and electroproduction off protons is sensitive to the contributions from most excited nucleon states. Together with N$\pi$ channels, the studies of  $\pi^+\pi^-p$ photo- and electroproduction represent the major source of the information on the N$^*$ photo-/electroexcitation amplitudes off protons, the so-called resonance photocouplings and $\gamma_{v}pN^*$-electrocouplings \cite{Bu12,BurkertRoberts,Mok18}.

The $\pi^+\pi^- p$ electroproduction data from CLAS~\cite{Ri03,Fe09,Is17,Fed18,Gol18} provided the only available information on nine independent single-differential and fully-integrated cross sections binned in $W$ and $Q^2$ in the mass range $W < 2.0$~GeV and at photon virtualities of 0.25~GeV$^2 < Q^2 < 5.0$~GeV$^2$. The kinematic coverage and the measured observables from this exclusive channel are shown in Table~\ref{tab-1} together with the observables from other exclusive meson electroproduction channels relevant in the resonance excitation region. The numerical results on the measured observables are available in the CLAS Physics Data Base \cite{clasdb}. These data were obtained with almost complete coverage of the final state phase space, which is of particular importance for extraction of the resonance parameters.   

\begin{table}[htb!]
\begin{center}

\begin{tabular}{|c|c|c|c|} \hline
Hadronic       &  $W$-range    & $Q^2$-range     & Measured observables \\
final state    &  GeV          & GeV$^2$         &     \\  \hline
$\pi^+ n$      &  1.10-1.38     & 0.16-0.36      & $\frac{d\sigma}{d\Omega}$ \\
               &  1.10-1.55     & 0.30-0.60      & $\frac{d\sigma}{d\Omega}$ \\
               &  1.10-1.70     & 1.70-4.50      & $\frac{d\sigma}{d\Omega}$, $A_{LT'}$ \\
               &  1.60-2.00     & 1.80-4.50      &  $\frac{d\sigma}{d\Omega}$    \\ \hline	       
$\pi^0 p$      &  1.10-1.38     & 0.16-0.36      & $\frac{d\sigma}{d\Omega}$ \\
               &  1.10-1.68     & 0.40-1.15      & $\frac{d\sigma}{d\Omega}$, $A_{LT'}$, $A_t$, $A_{et}$ \\
               &  1.10-1.39     & 3.00-6.00      & $\frac{d\sigma}{d\Omega}$  \\ \hline     
$\eta p$       &  1.50-2.30     & 0.20-3.10      & $\frac{d\sigma}{d\Omega}$ \\ \hline     
$K^+\Lambda$   &  1.62-2.60     & 1.40-3.90      & $\frac{d\sigma}{d\Omega}$ \\
               &  1.62-2.60     & 0.70-5.40      & $P'$, $P^0$ \\ \hline     
$K^+\Sigma^0$   &  1.62-2.60     & 1.40-3.90     & $\frac{d\sigma}{d\Omega}$ \\
                &  1.62-2.60     & 0.70-5.40     & $P'$ \\ \hline     
$\pi^+\pi^-p$   &  1.30-1.60     & 0.20-0.60      & Nine single-differential \\
                &  1.40-2.10     & 0.50-1.50      & cross sections \\
                &  1.40-2.00     & 2.00-5.00      &                 \\ 
                &  1.30-1.83     & 0.40-1.00      &                 \\ 
                &  1.60-2.50     &   0      &                 \\ \hline
\end{tabular}
\caption{\label{tab-1} Observables for exclusive meson electroproduction off protons that have been 
measured with the CLAS detector in the resonance region and stored in the CLAS Physics Database 
\cite{clasdb}: center-of-mass (CM) angular distributions for the final mesons ($\frac{d\sigma}{d\Omega}$); beam, target, and 
beam-target asymmetries ($A_{LT'}$, $A_t$, $A_{et}$); and recoil hyperon polarizations ($P'$, $P^0$).}
\end{center}
\end{table}

\begin{figure*}[htp]
\begin{center}
\includegraphics[width=6cm]{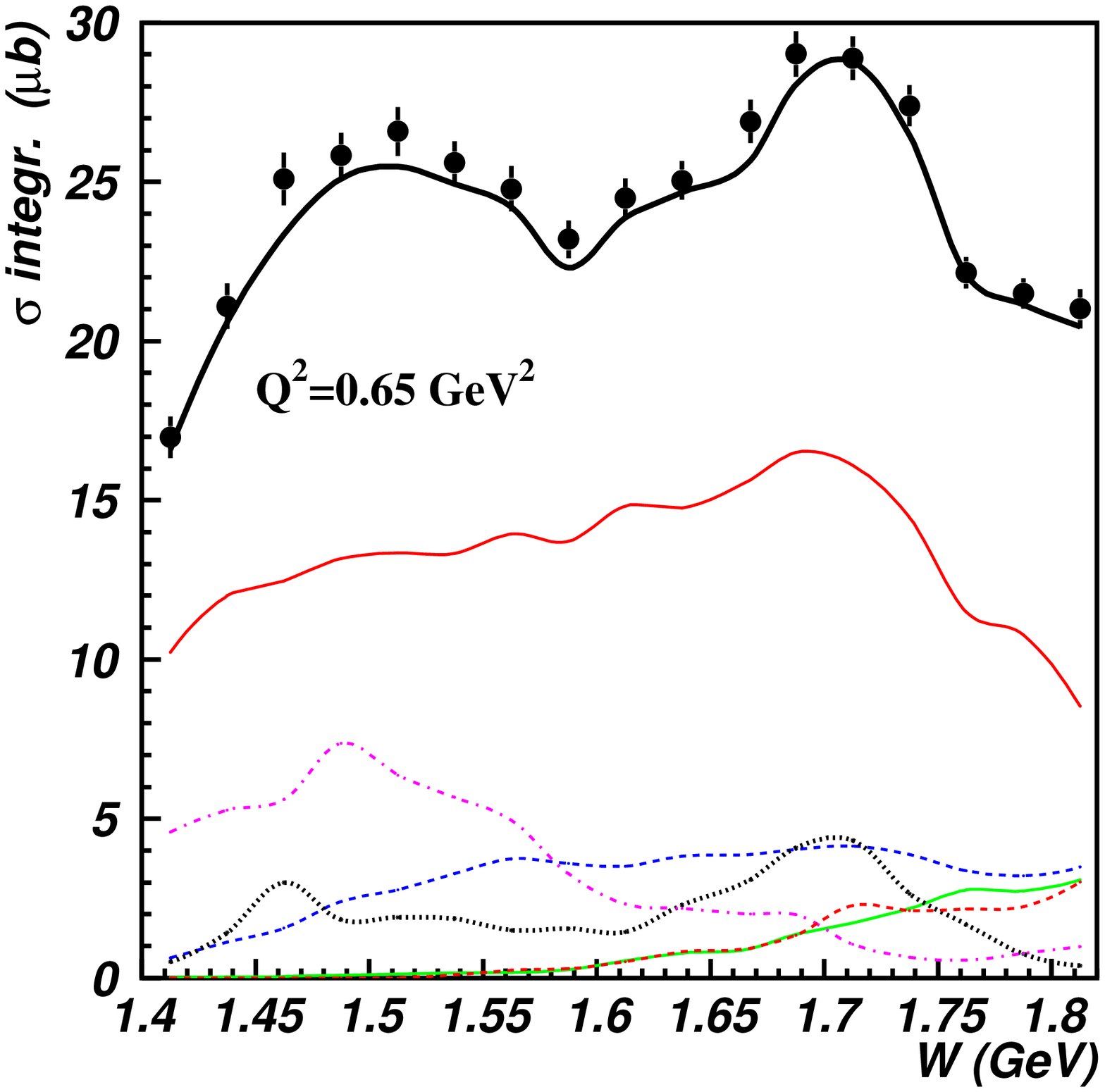}
\includegraphics[width=6cm]{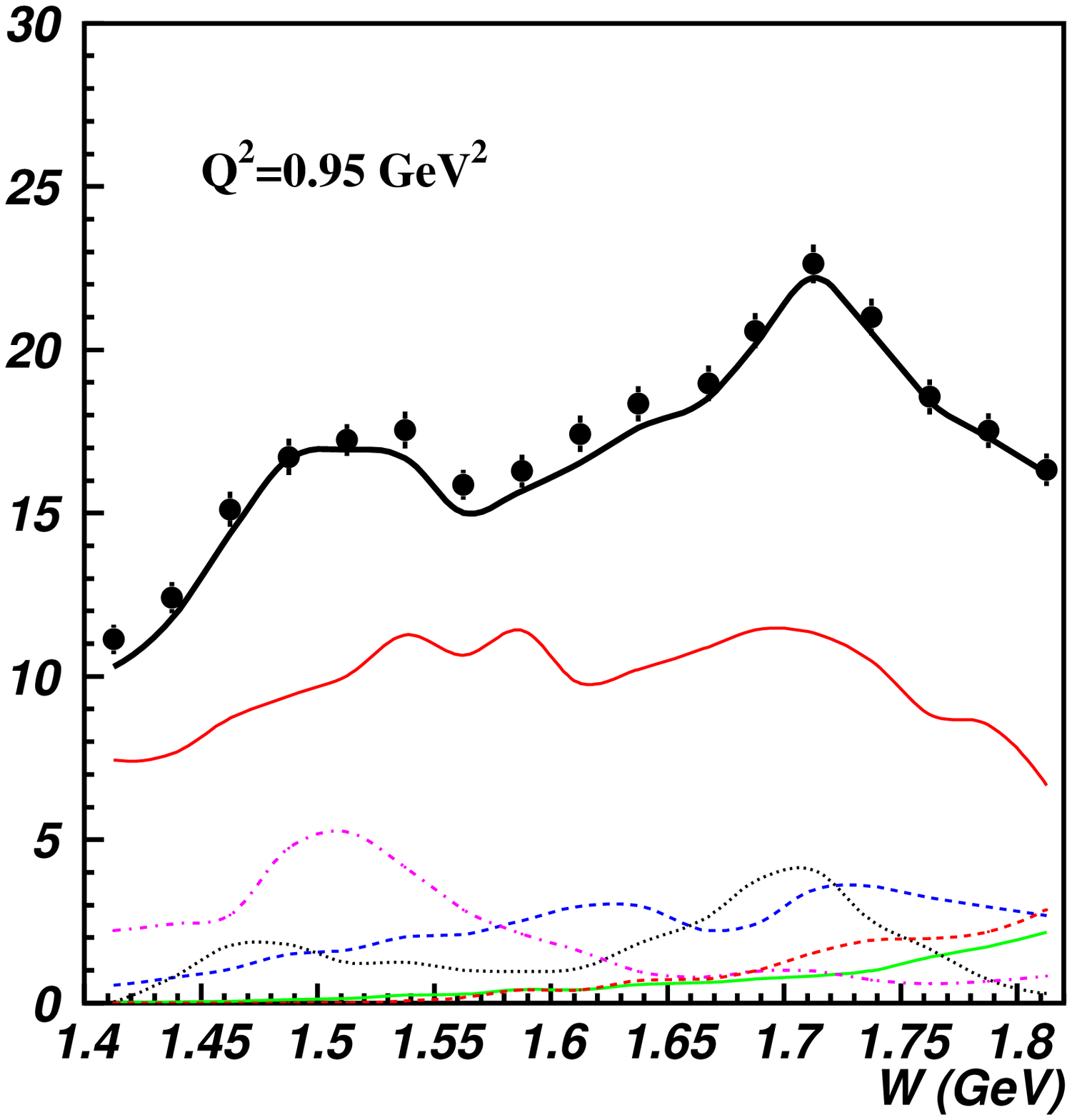}
\includegraphics[width=6cm]{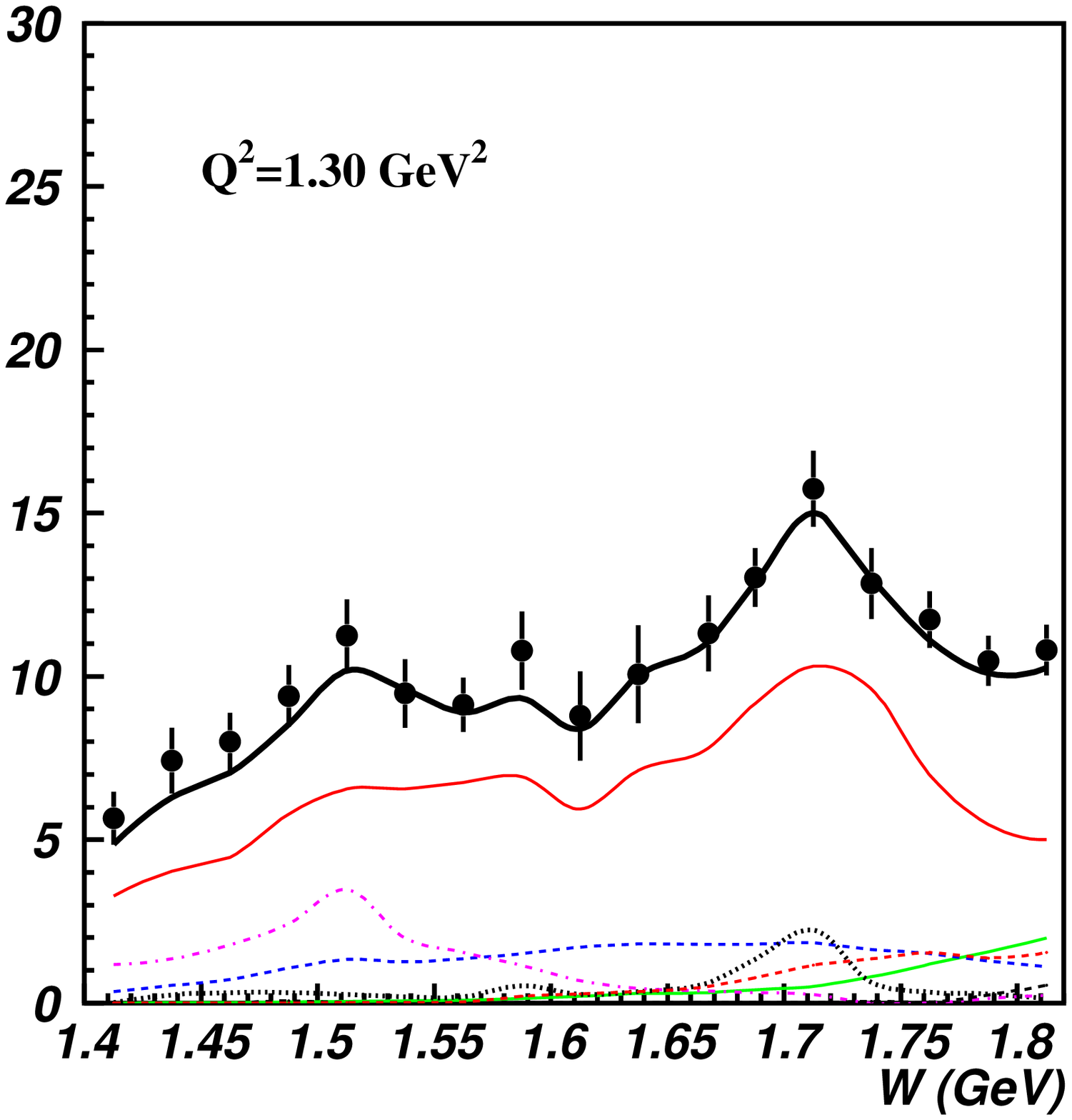}
\vspace{-0.1cm}
\caption{(Color Online) Description of the fully integrated $\pi^+\pi^-p$ electroproduction cross sections 
achieved within the framework of the JM model together with the 
cross sections for the various contributing mechanisms: full cross section (black solid), $\pi^-\Delta^{++}$ 
(red thin solid), $\rho p$ (green thin solid), $\pi^+\Delta^0$ (blue thin dashed), $\pi^+ N^0(1520)3/2^-$ (black 
dotted), direct 2$\pi$ mechanisms (magenta  thin dot-dashed), and $\pi^+ N^0(1685)5/2^+$ (red thin dashed).}  
\label{integsec}
\end{center}
\end{figure*}

\begin{figure*}[htp]
\begin{center}
\includegraphics[width=8.5cm]{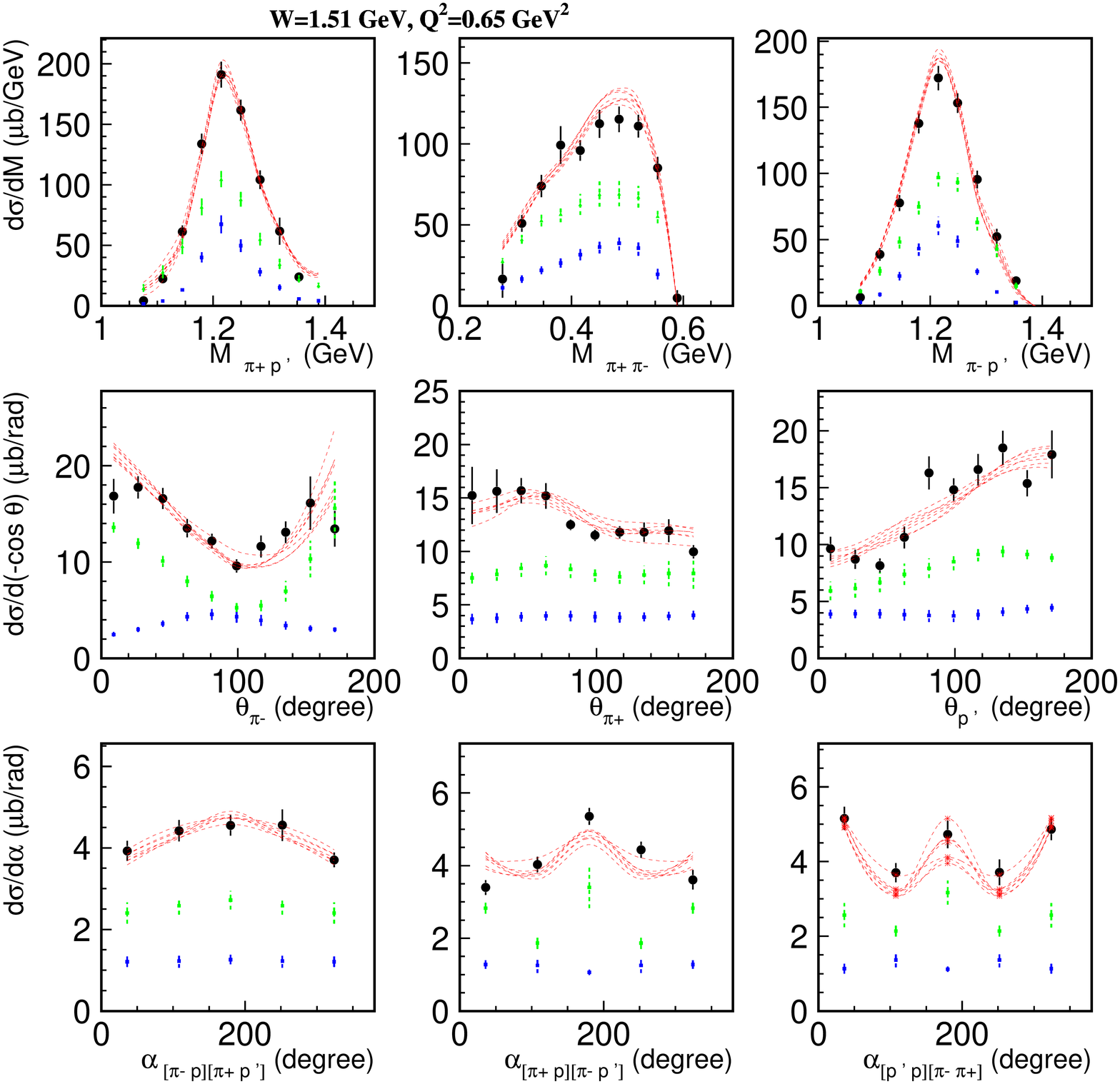}
\includegraphics[width=8.5cm]{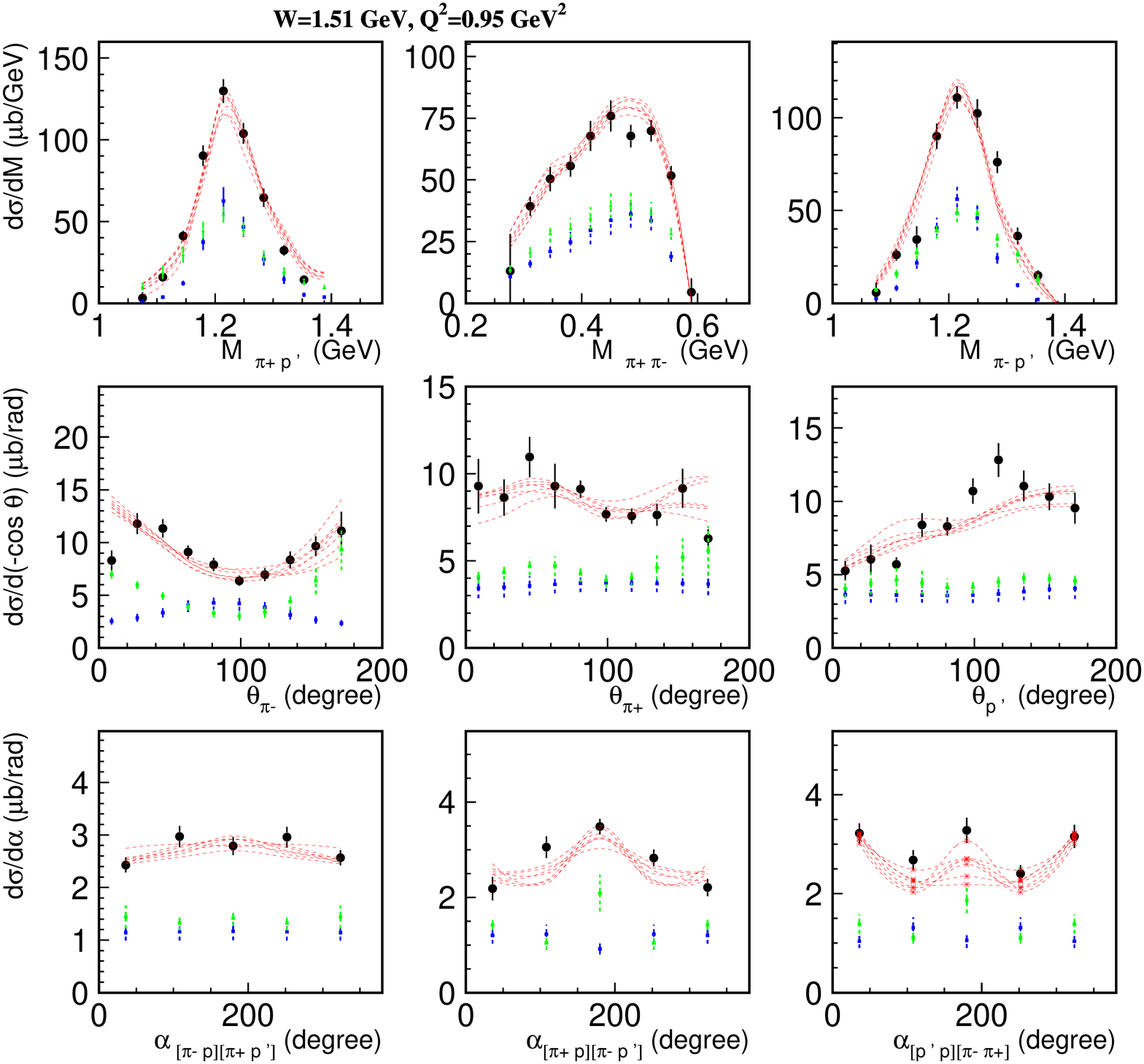}\\
\vspace{-0.1cm}
\caption{(Color Online) Examples of fits to the CLAS data~\cite{Ri03} on the nine one-fold differential 
$\pi^+\pi^-p$ electroproduction cross sections in particular bins of $W$ and $Q^2$ within the framework of the 
JM model. The curves correspond to those fits with $\chi^2/d.p.$ 
less then the value determined so that the selected in the fit cross sections are spread within the data uncertainties for the dominant part of the data points. The resonant and non-resonant contributions determined 
from the data fit within the framework of the JM model~\cite{Mo16} are shown by blue triangles and green squares, 
respectively.} 
\label{fitsec}
\end{center}
\end{figure*}

\begin{figure*}[htp]
\begin{center}
\includegraphics[width=8.5cm]{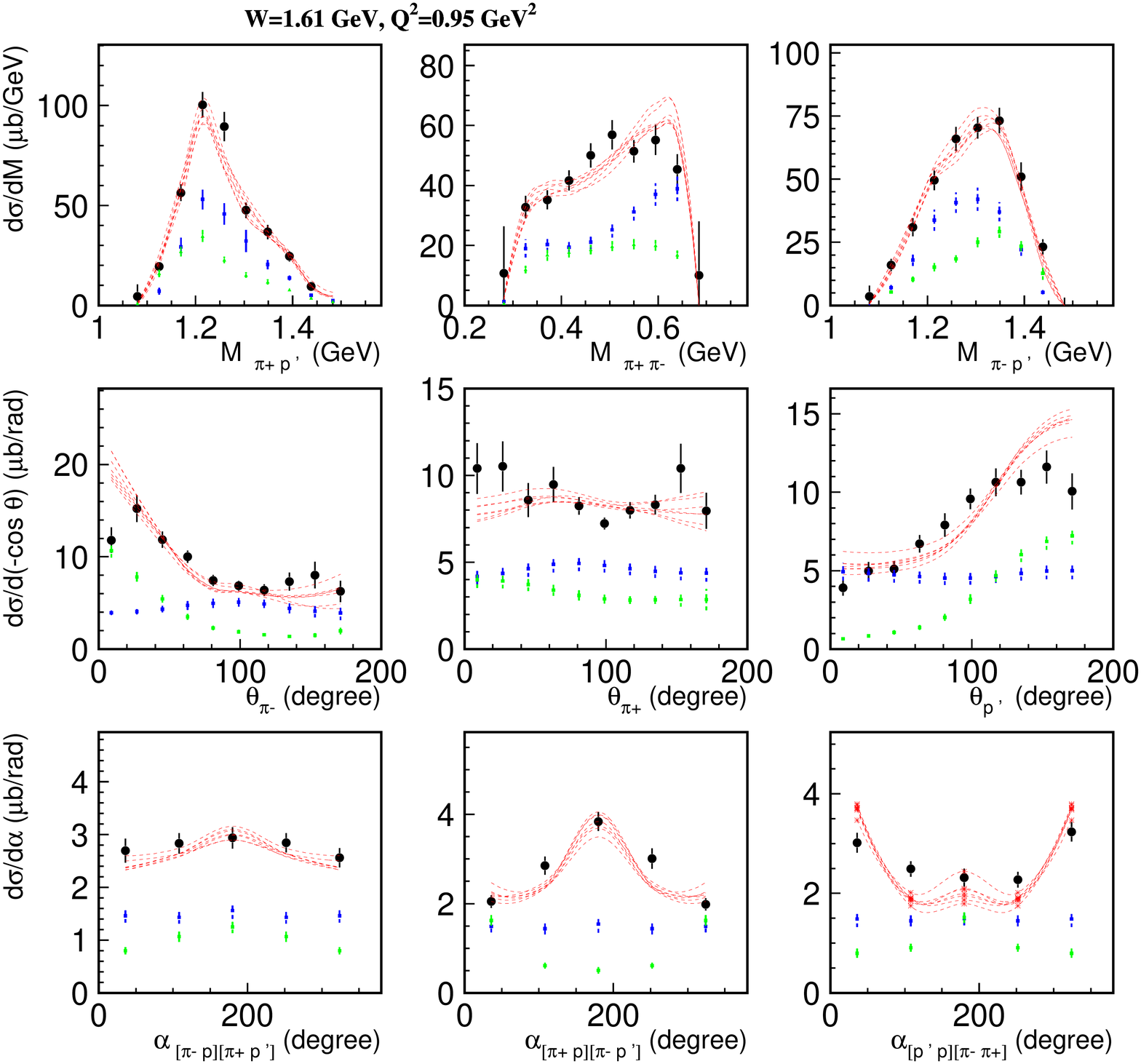}
\includegraphics[width=8.5cm]{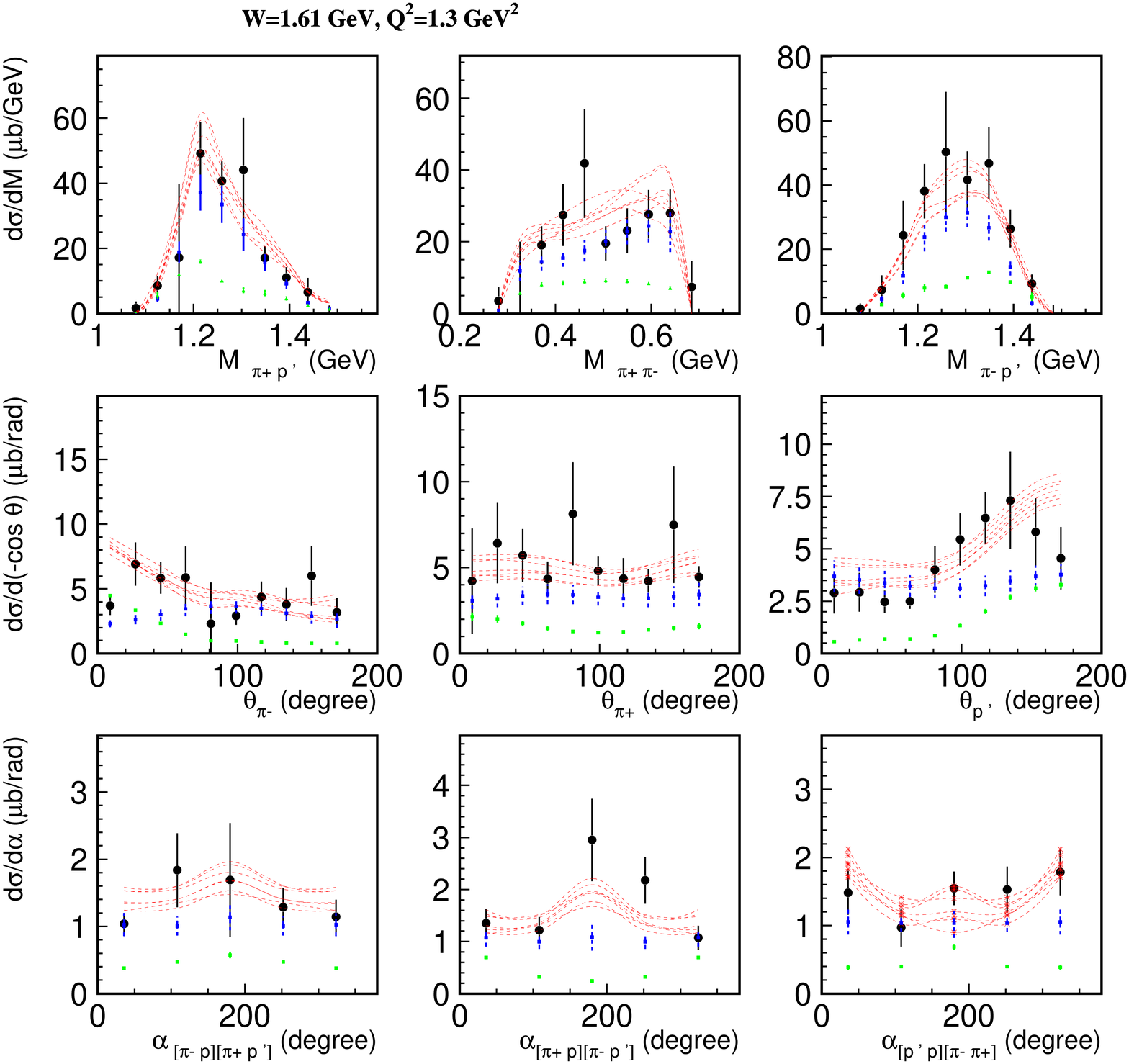}
\vspace{-0.1cm}
\caption{(Color Online) The same as in Fig.~\ref{fitsec}, but in other bins of $W$ and $Q^2$.} 
\label{fitsec1}
\end{center}
\end{figure*}

The analysis of $\pi^+\pi^-p$ electroproduction off proton data from CLAS has allowed us to develop the Jefferson Lab-Moscow State University (JM) reaction model~\cite{Gol18,Mo09,Mo12,Mo16} with the goal to extract resonance photo-/electrocouplings, as well as the $\pi\Delta$ and $\rho p$ hadronic decay widths. This model incorporates all relevant reaction mechanisms in the $\pi^+\pi^-p$ final-state channel that contribute significantly to the measured photo-/electroproduction cross sections off protons in the resonance region, including the 
$\pi^-\Delta^{++}$, $\pi^+\Delta^0$, $\rho^0 p$, $\pi^+N^{0}(1520)3/2^-$, $\pi^+N^{0}(1685)5/2^+$ meson-baryon channels, as well as the direct production of the $\pi^+\pi^-p$ final state without formation of intermediate unstable hadrons. The contributions from well established $N^*$ states in the mass range up to 2.0~GeV were included into 
the amplitudes of the $\pi\Delta$ and $\rho p$ meson-baryon channels by employing a unitarized version of the Breit-Wigner ansatz~\cite{Mo12}. The JM model provides a good description of the $\pi^+\pi^- p$ differential cross sections at $W < 2.0$~GeV and 0.2~GeV$^2 < Q^2 < 5.0$~GeV$^2$ with $\chi^2/d.p. < 3.0$ accounting for only the statistical uncertainties of
the data. Descriptions of the fully integrated $\pi^+\pi^-p$ electroproduction cross sections are shown in 
Fig.~\ref{integsec} together with the contributions from the meson-baryon mechanisms of the JM model inferred from the CLAS data. These results offer valuable input for the global multi-channel analyses in the resonance excitation region within advanced coupled channel approaches~\cite{Lee10,Lee101,Do18} in particular for their extensions towards the extraction of $\gamma_{v}pN^*$-electrocouplings from exclusive meson electroproduction data. The quality of the description of the CLAS data suggests the reliable separation of resonant and non-resonant contributions achieved by fitting the CLAS data. The representative examples for the resonant/non-resonant contributions to the 
nine one-fold differential cross sections determined from the CLAS data are shown in Fig.~\ref{fitsec},~\ref{fitsec1}. The resonant/non-resonant contributions are located within well 
defined ranges (see Fig.~\ref{fitsec},~\ref{fitsec1}) and show no evidence for separation ambiguities. 
Furthermore, the uncertainties of the resonant/non-resonant contributions are comparable with the 
uncertainties of the measured cross sections, demonstrating again unambiguous resonant/non-resonant separation of 
a good accuracy. The isolation of the resonant contributions makes it possible to determine the resonance photo-/electrocouplings along with the $\pi \Delta$ and $\rho N$ decay widths by employing the unitarized Breit-Wigner ansatz~\cite{Mo12} that fully accounts for the unitarity restrictions on the resonant amplitudes.

\begin{figure}[htb!]
\centering
\includegraphics[width=6cm,clip]{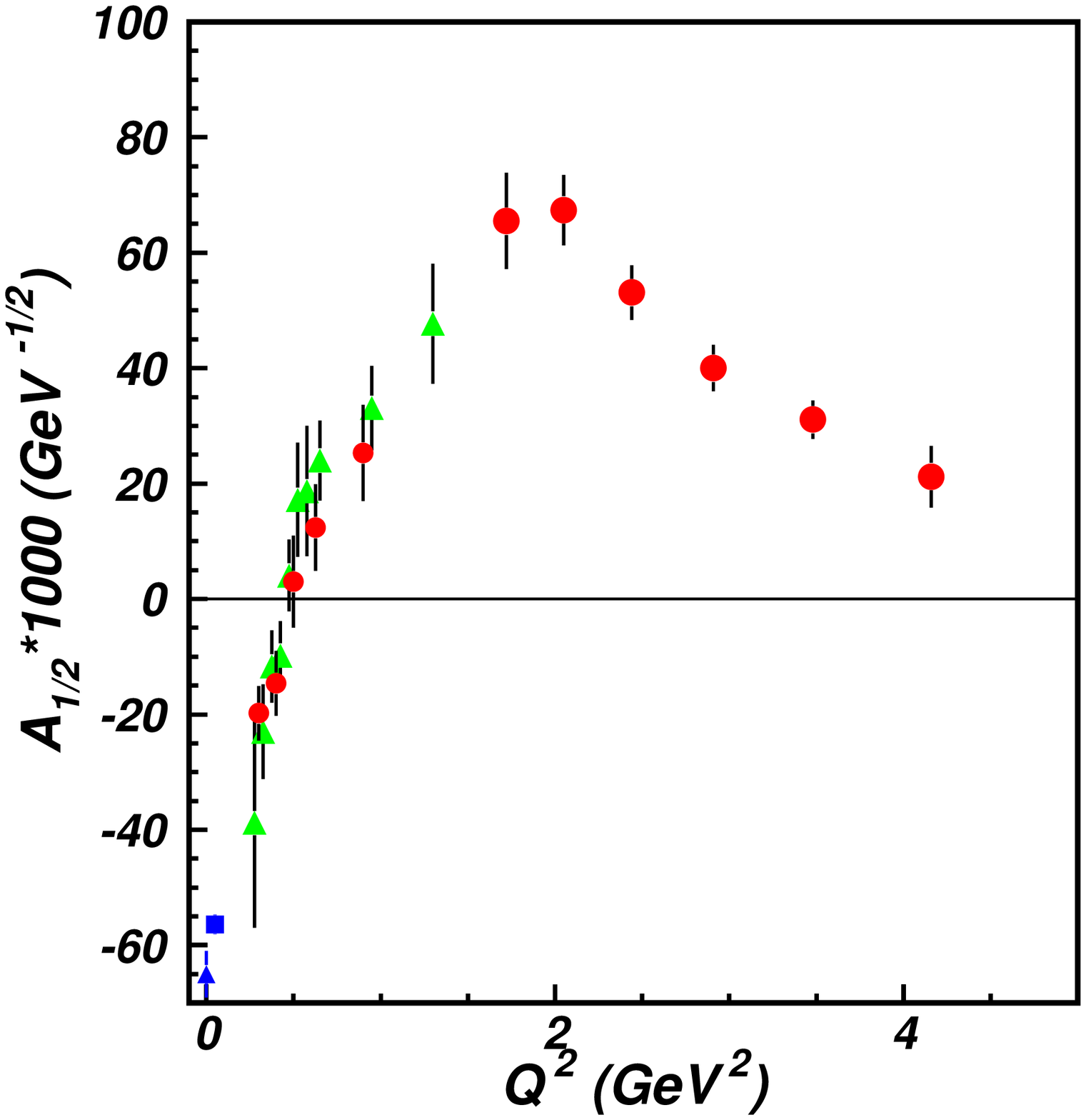}
\includegraphics[width=6cm,clip]{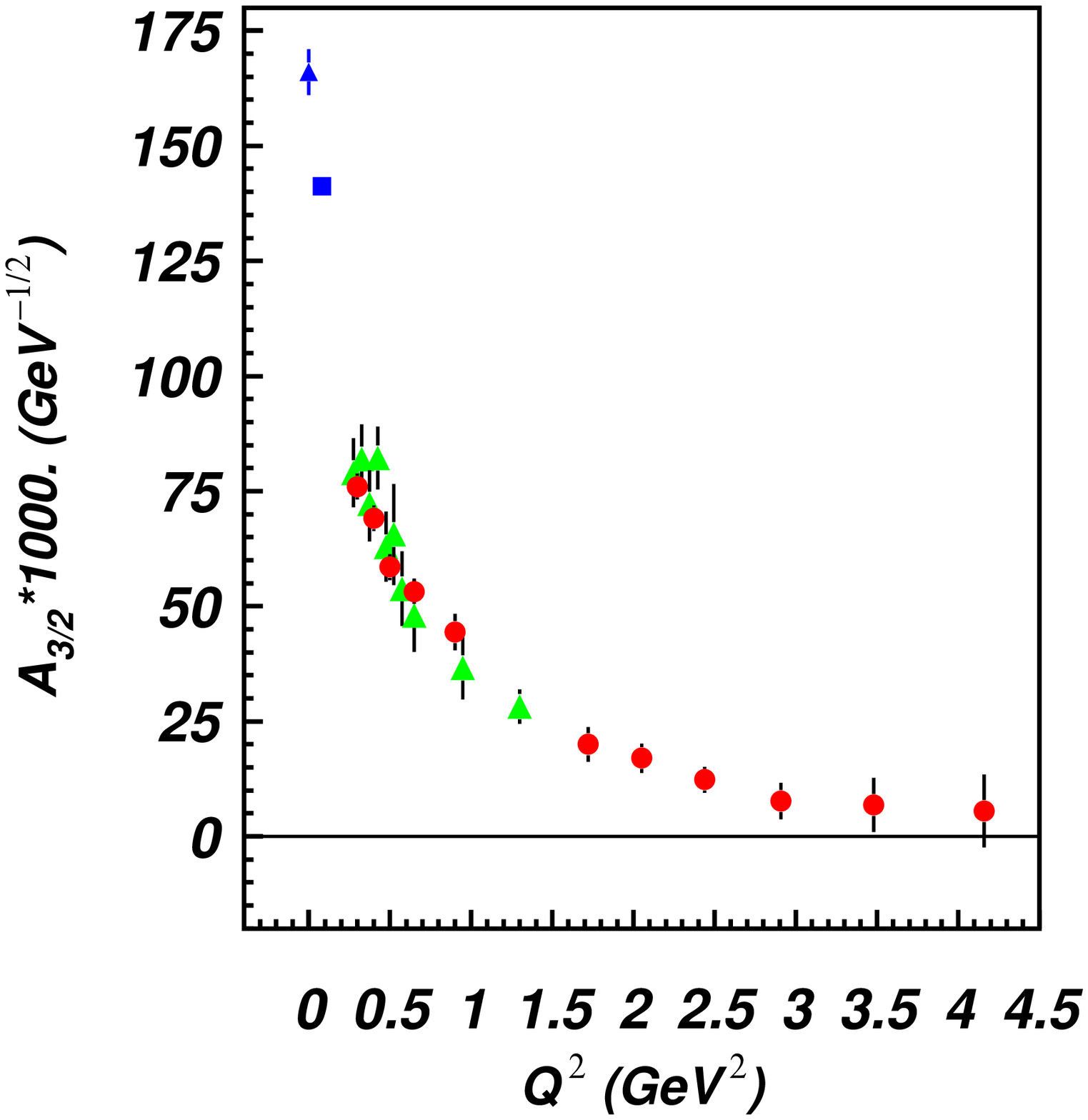}
\includegraphics[width=6cm,clip]{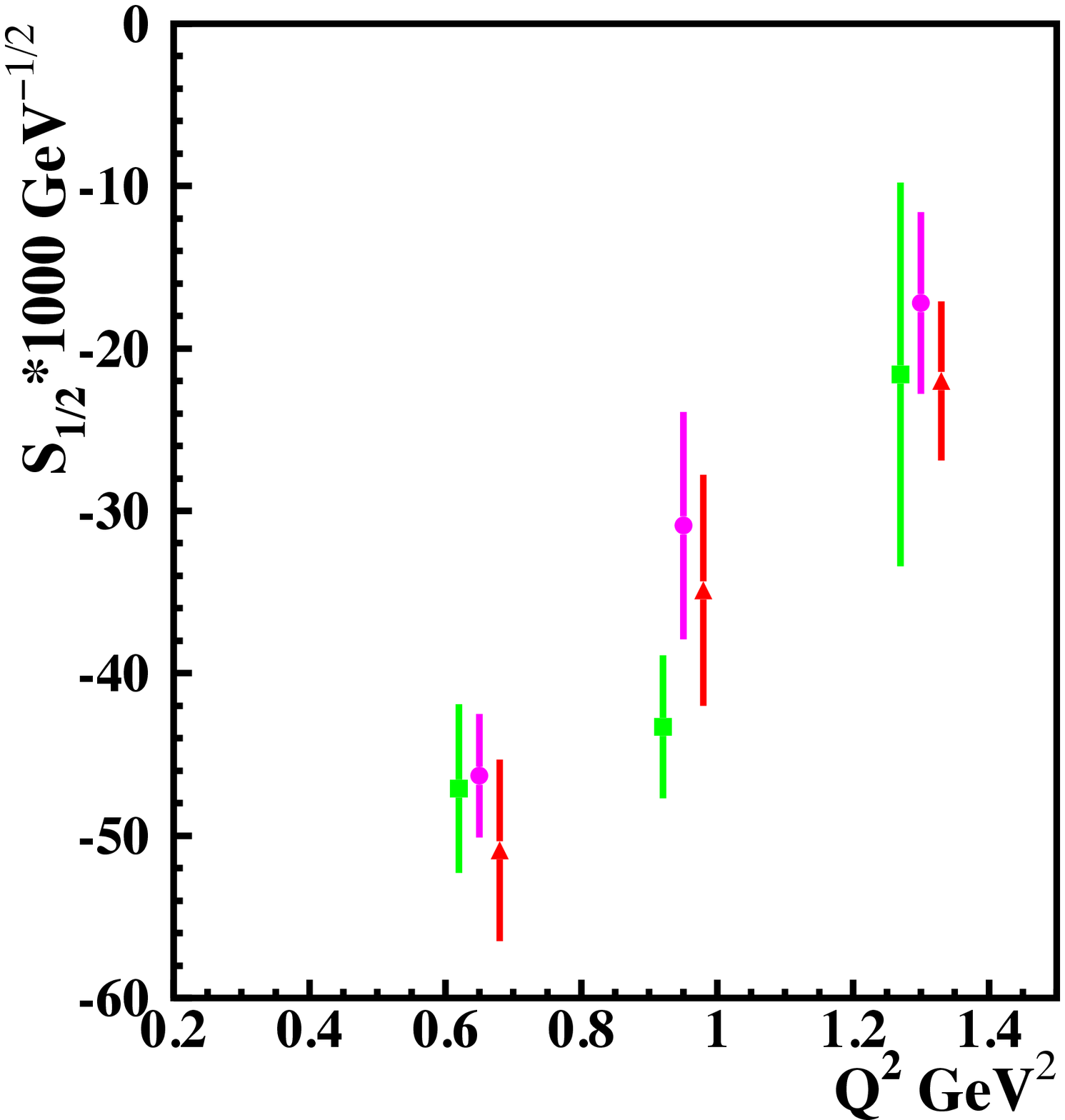} 
\caption{(Color Online) $A_{1/2}$ $\gamma_vpN^*$ electrocouplings of the $N(1440)1/2^+$ (top left), $A_{3/2}$ 
$\gamma_vpN^*$ electrocouplings of the $N(1520)3/2^-$ (top right), and  $S_{1/2}$ $\gamma_vpN^*$ electrocouplings of the $\Delta(1620)1/2^-$ (bottom) from analyses of the CLAS electroproduction data off protons in the $\pi N$~\cite{Az09} 
(red circles in the left and middle panels) and $\pi^+\pi^-p$ channels~\cite{Mo12,Mo16} (green triangles in the left and middle 
panels). The right panel shows the $\Delta(1620)1/2^-$ electrocouplings obtained from analyses of $\pi^+\pi^-p$ electroproduction data off protons \cite{Mo16} carried out independently in three intervals of $W$: 1.51~GeV $\to$ 1.61~GeV (green squares), 1.56~GeV $\to$ 1.66~GeV (magenta circles), and 1.61~GeV $\to$ 1.71~GeV (red triangles). The photocouplings were taken from the RPP~\cite{rpp} (blue filled triangles) and the CLAS data analysis~\cite{Dug09} of $\pi N$ photoproduction (blue filled squares). }
\label{p11d13s31}
\end{figure}

\subsection{The resonance parameters from the CLAS $\pi^+\pi^-p$ electroproduction off protons data}
Table~\ref{tab-2} summarizes the available CLAS results on the $\gamma_vpN^*$ electrocouplings from $\pi^+\pi^-p$ electroproduction off protons and from other exclusive meson electroproduction channels. The numerical values of these electrocouplings from CLAS can be found in Ref.~\cite{mokeev-web}. The computer code for interpolation/extrapolation over $Q^2$ in the range of $Q^2$ up to 5.0 GeV$^2$ of the CLAS results on resonance electrocouplings is available on the web page of Ref.~\cite{isupov-web}. The CLAS data on exclusive charged double pion electroproduction off protons provided information on electroexcitation amplitudes of most excited nucleon states in the mass range up to 1.8 GeV and at photon virtualities $Q^2$ $<$ 1.5 GeV$^2$. The results on the $\gamma_vpN^*$ electrocouplings of excited nucleon states in the mass range W $<$ 1.6 GeV have become available from independent studies of N$\pi$, N$\eta$, and $\pi^+\pi^-p$ electroproduction off protons. 

\begin{table}[htb!]
\begin{center}
\begin{tabular}{|c|c|c|} \hline
Exclusive        &  Excited proton    & Coverage over $Q^2$ for extracted    \\
channel          &  state             & $\gamma_{v}pN^*$ electrocouplings, GeV$^2$    \\ \hline
$\pi^+ n$, $\pi^0 p$       &  $\Delta(1232)3/2^+$,                                & 0.16\textemdash 6.00       \\
                           &  $N(1440)1/2^+$, $N(1520)3/2^-$, $N(1535)1/2^-$      & 0.30\textemdash 4.16       \\ \hline	       
$\pi^+ n$      &  $N(1675)5/2^-$, $N(1680)5/2^+$, $N(1710)1/2^+$                  & 1.60\textemdash 4.50        \\ \hline     
$\eta p$       &  $N(1535)1/2^-$        & 0.20\textemdash 2.90       \\ \hline     
$\pi^+\pi^-p$   & $N(1440)1/2^+$, $N(1520)3/2^-$       & 0.25\textemdash  1.50       \\
                &  $\Delta(1620)1/2^-$, $N(1650)1/2^-$, $N(1680)5/2^+$      & 0.50\textemdash 1.50       \\
                &  $\Delta(1700)3/2^-$, $N(1720)3/2^+$, $N'(1720)3/2^+$     &         \\ \hline
\end{tabular}
\caption{\label{tab-2} Resonance electrocouplings available from the analyses of the CLAS data on exclusive meson electroproduction off protons in the resonance region.}
\end{center}
\end{table}

Consistent results for the $\gamma_vpN^*$ electrocouplings of the $N(1440)1/2^+$ and $N(1520)3/2^-$ resonances, which have been determined in independent analyses of the dominant meson electroproduction channels $\pi N$ and $\pi^+\pi^-p$, shown in Fig.~\ref{p11d13s31} (left) and (middle), demonstrate that the extraction of these fundamental quantities is reliable. Studies of the exclusive electroproduction channels off protons $\pi N$ and $\pi^+\pi^-p$ offer complementary information on the $N^*$ electrocouplings. For low-lying excited nucleon states in the mass range up to 1.6~GeV that decay preferentially to the $\pi N$ final states, the data on single-pion exclusive electroproduction drive the extraction of these resonance electrocouplings. However, studies of the $\eta p$ and $\pi^+\pi^-p$ channels are needed in order to validate the $\gamma_vpN^*$ electrocoupling extraction from $\pi N$ electroproduction. Although the non-resonant contributions in different exclusive channels are entirely different, the resonance electrocouplings extracted from different exclusive channels should be the same. Consistent results on the $\gamma_vpN^*$ electrocouplings obtained from independent analyses of different exclusive channels  validate the extraction of these quantities.

The CLAS data for the $\pi^+ \pi^- p$ channel play a critical role in the extraction of the $\gamma_vpN^*$ electrocouplings of  higher-lying nucleon excited states ($M >1.60$~GeV), which decay preferentially to the $\pi\pi N$ final states, e.g. $\Delta(1620)1/2^-$,  $\Delta(1700)3/2^-$, $N(1720)3/2^+$, and the $N'(1720)3/2^+$ candidate state. Right now, the electrocouplings of these states can only be determined from the data in the $\pi^+\pi^-p$ exclusive electroproduction channel off protons, while the $\pi N$ channels do not have enough sensitivity to the electrocouplings of the aforementioned resonances.

We have developed special procedures to test the reliability of the $\gamma_vpN^*$ resonance electrocouplings extracted from the charged double pion electroproduction data. In this case, we carried out the extraction of the resonance parameters, independently fitting the CLAS $\pi^+\pi^-p$ electroproduction data~\cite{Ri03} in overlapping intervals of $W$. The non-resonant amplitudes in each of the $W$-intervals are different, while the resonance parameters should remain the same as 
they are determined from the data fit in different $W$-intervals, see Fig.~\ref{p11d13s31} (right). The consistent results on 
these electrocouplings from the independent analyses in different $W$-intervals strongly support their reliable extraction. The tests described above demonstrated the capability of the models   to provide reliable information on the $\gamma_vpN^*$ resonance electrocouplings from independent analyses of the data on exclusive $\pi N$ and $\pi^+\pi^-p$ electroproduction.

\begin{table*}
\begin {center} 
 \footnotesize

\begin{tabular}{ c  
                 D{,}{\pm}{-1}  
		 D{,}{\hspace{3pt} \text{--} \hspace{3pt}}{-1} 
                 D{,}{\pm}{-1}   
		 D{,}{\pm}{-1} 
		 D{,}{\hspace{3pt} \text{--} \hspace{3pt}}{-1}   
		 D{,}{\pm}{-1}} \hline
\toprule
{} 
& \multicolumn{1}{l}{$A_{1/2}\times 10^3$}
& \multicolumn{1}{l}{$A_{1/2}\times 10^3$}
& \multicolumn{1}{l}{$A_{1/2}\times 10^3$}
& \multicolumn{1}{l}{$A_{3/2}\times 10^3$}
& \multicolumn{1}{l}{$A_{3/2}\times 10^3$}
& \multicolumn{1}{l}{$A_{3/2}\times 10^3$}
\\  \hline
{Resonances}
& \multicolumn{1}{l}{from $\pi^+\pi^-p$}
& \multicolumn{1}{l}{PDG ranges}
& \multicolumn{1}{l}{multichannel}
& \multicolumn{1}{l}{from $\pi^+\pi^-p$}
& \multicolumn{1}{l}{PDG ranges}
& \multicolumn{1}{l}{multichannel}
\\
{}
& \multicolumn{1}{l}{GeV$^{-1/2}$}
& \multicolumn{1}{l}{GeV$^{-1/2}$}
& \multicolumn{1}{l}{analysis~}
& \multicolumn{1}{l}{GeV$^{-1/2}$}
& \multicolumn{1}{l}{GeV$^{-1/2}$}
& \multicolumn{1}{l}{analysis~}
\\
& \multicolumn{1}{l}{}
& \multicolumn{1}{l}{}
& \multicolumn{1}{l}{GeV$^{-1/2}$}
& \multicolumn{1}{l}{}
& \multicolumn{1}{l}{}
& \multicolumn{1}{l}{GeV$^{-1/2}$}
\\ \hline
\midrule
$\Delta(1620)1/2^-$    & 29.0,6.2    &	30 , 60     &   55,7      & 	        &  	        &	     \\
$N(1650)1/2^-$         & 60.5,7.7    &	35 , 55     &   32,6      & 	        &  	        &	     \\
$N(1680)5/2^+$         & -27.8,3.6   &  -18 ,{-5}   &  -15,2      & 128,11      &  130 , 140    &  136 ,  5  \\
$N(1720)3/2^+$         & 80.9,11.5   &  80 , 120    &   115,45    & -34.0,7.6   &  -48 , 135    &  135 , 40  \\
$\Delta(1700)3/2^-$    & 87.2,18.9   &  100 , 160   &   165,20    & 87.2,16.4   &   90 , 170    &  170 , 25  \\
$\Delta(1905)5/2^+$    & 19.0,7.6    &  17 , 27     &   25,5      & -43.2,17.3  &  -55 , {-35}  &  -50  , 5  \\
$\Delta(1950)7/2^+$    & -69.8,14.1  &  -75 , {-65} &   -67,5     & -118.1,19.3 &  -100 , {-80} &  -94 , 4   \\  \hline
\bottomrule	        	      
\end{tabular}
\normalsize
\caption{Resonance photocouplings determined from analysis of the $\pi^+\pi^-p$ photoproduction data from \cite{Gol18} in 
comparison with the previous results from the PDG average and from multichannel analysis \cite{So17}.}
\label{nstpar}
\end {center}
\end{table*}

Recently, information on the excited nucleon state photocouplings was extended by the first results on the photocouplings of most nucleon resonances with masses above 1.6 GeV from exclusive $\pi^+\pi^-p$ photoproduction off protons measured with CLAS \cite{Gol18}. Many of them decay preferentially to the $\pi\pi N$ final states. 
The cross sections of $\pi^+\pi^-p$ photoproduction off protons are the biggest among $\pi\pi N$ photoproduction off proton channels. Overall, $\approx$400 million $\pi^+\pi^-p$ events 
were selected exceeding by a factor of $\sim$50 the statistics previously collected in this channel. Good description of these new data with $1.15 < \chi^2/d.p. < 1.3$ was achieved within the framework of the meson-baryon reaction model JM accounting for the contributions from the statistical and systematical data uncertainties, combined. 
Credible isolation of the resonant contributions obtained in the data fit allowed us to determine the resonance parameters. 
The resonance photocouplings extracted from this work are listed in Table~\ref{nstpar} and compared with the resonance
photocoupling ranges and the results of the multichannel analysis \cite{So17} included in the PDG2018 \cite{rpp}. There is good agreement in the magnitude and sign of the photocouplings between our results and the photocoupling
ranges in the PDG listings. On the other hand, for several resonances, the photocouplings determined
from the multichannel analysis are different from ours. Implementation of our $\pi^+\pi^-p$
photoproduction data into the global multichannel analyses will improve essentially our knowledge on the photocouplings and hadronic decay parameters 
of the resonances in the mass range of W $>$ 1.6 GeV. For a successful description of the  $\pi^+\pi^-p$ photo- and electroproduction CLAS data with $Q^2$-independent resonance hadronic decay parameters, 
the new baryon states N'(1720)3/2$^+$ is needed. Evidences for this new state will be published soon.

\section{Extending knowledge on the nucleon resonance structure and strong QCD}
Experimental results on the $\gamma_vpN^*$ electrocouplings offer insight into excited nucleon structure and the 
dynamics of strong QCD.

\begin{figure}[htb!]
\centering
\includegraphics[width=7.5cm,clip]{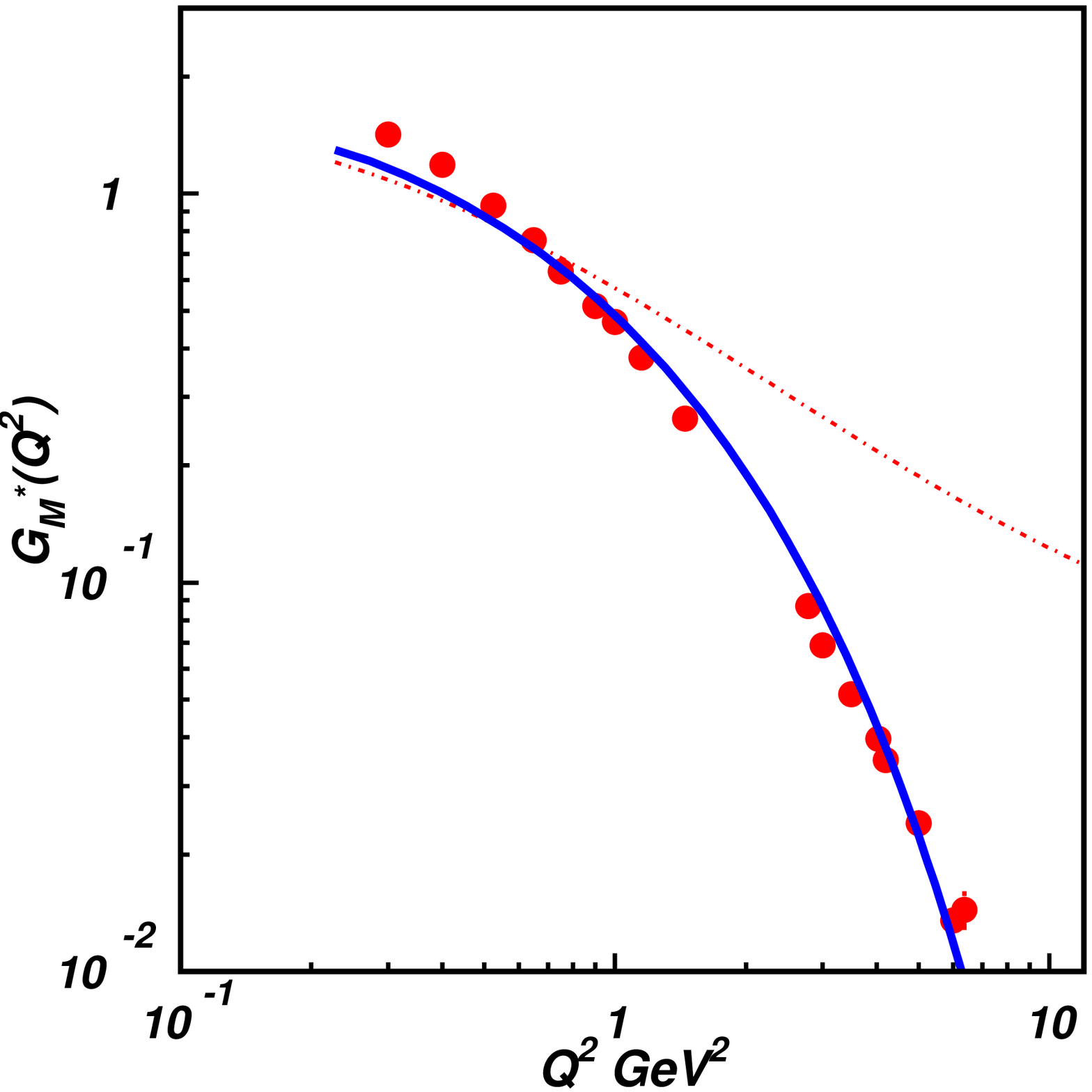}
\includegraphics[width=6.5cm,clip]{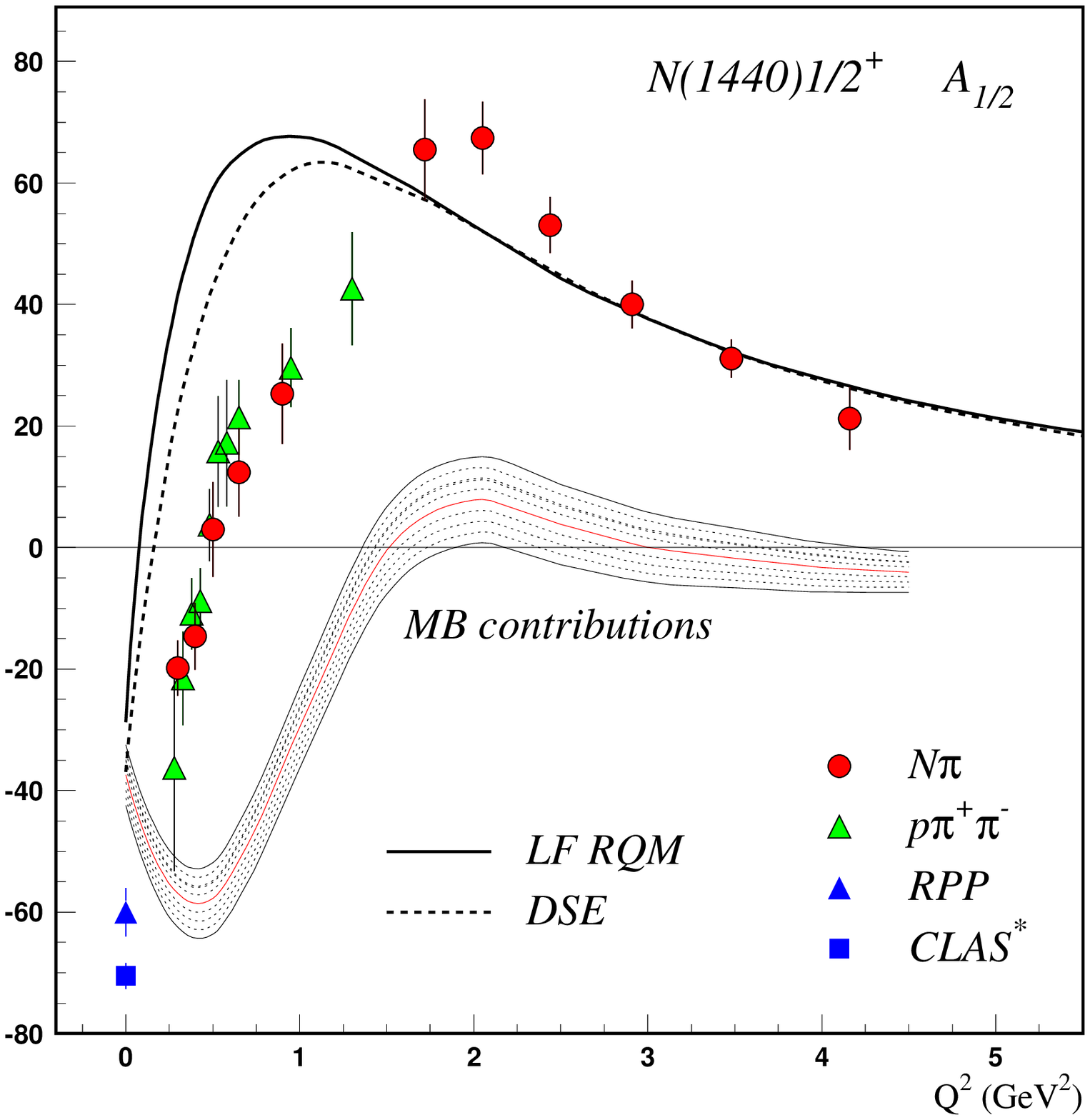}
\caption{(Color Online) Insight into the resonance structure and strong QCD dynamics from the experimental results on the 
magnetic $p \to \Delta(1232)3/2^+$ transition form factor (left) \cite{Az09} and the $A_{1/2}$ electrocoupling of the $N(1440)1/2^+$ resonance (right)~\cite{Az09,Mo12,Mo16}. The red dashed and blue solid curves in the left panel correspond 
to the computation of the magnetic $p \to \Delta(1232)3/2^+$ transition form factor starting from the QCD-Lagrangian within DSE by employing a simplified contact $qq$-interaction (frozen quark mass) and a realistic $qq$-interaction (running quark mass), respectively \cite{Seg14}. In the right panel the DSE computation of the $A_{1/2}$ electrocoupling of the $N(1440)1/2^+$ resonance with a realistic $qq$-interaction and the same running quark mass as employed in the successful evaluations of nucleon elastic and magnetic $p \to \Delta(1232)3/2^+$ form factors is shown by the dashed line~\cite{Seg15}. The result from a novel light front quark model \cite{Az12,Az17}, which incorporates a momentum-dependent dressed quark mass, is shown by the solid line. The shadowed area represents the meson-baryon cloud contribution inferred from the experimental results on the $A_{1/2}$ electrocoupling of the $N(1440)1/2^+$ resonance and the DSE evaluation \cite{Seg15} of the quark core contribution.}
\label{del33p11}
\end{figure}

Due to the rapid progress in the field of continuum QCD Dyson-Schwinger Equation (DSE) studies of excited nucleon states~\cite{Ro17,Seg14,Seg15}, the first evaluations of the magnetic $p \to \Delta(1232)3/2^+$ form factors and the $N(1440)1/2^+$ resonance electrocouplings 
starting from the QCD Lagrangian have recently become available. The $p \to \Delta(1232)3/2^+$ magnetic form factor and $A_{1/2}$ electrocoupling of the $N(1440)1/2^+$ resonance computed in Refs.~\cite{Seg14,Seg15} are shown in Fig.~\ref{del33p11}. These evaluations are applicable at photon virtualities  where the contributions of the inner quark core to the resonance electrocouplings are much larger than those from the external meson baryon cloud. In this range of photon 
virtualities, the evaluations~\cite{Seg14,Seg15} offer a good description of the experimental results on the 
$p \to \Delta(1232)3/2^+$ transition form factors and the $N(1440)1/2^+$ resonance electrocouplings. 

\begin{figure}[htb!]
\centering
\includegraphics[width=6cm,clip]{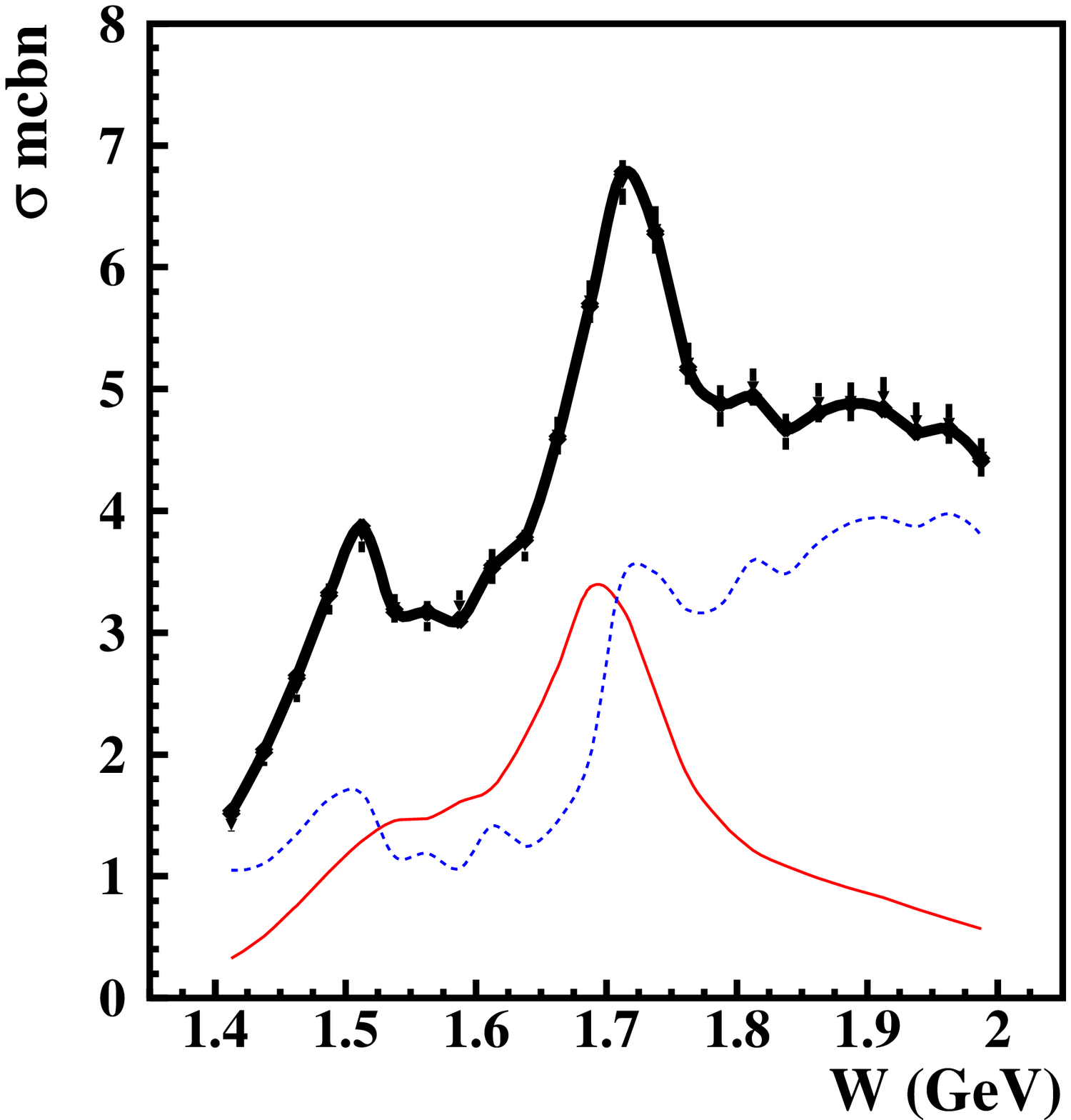}
\includegraphics[width=6cm,clip]{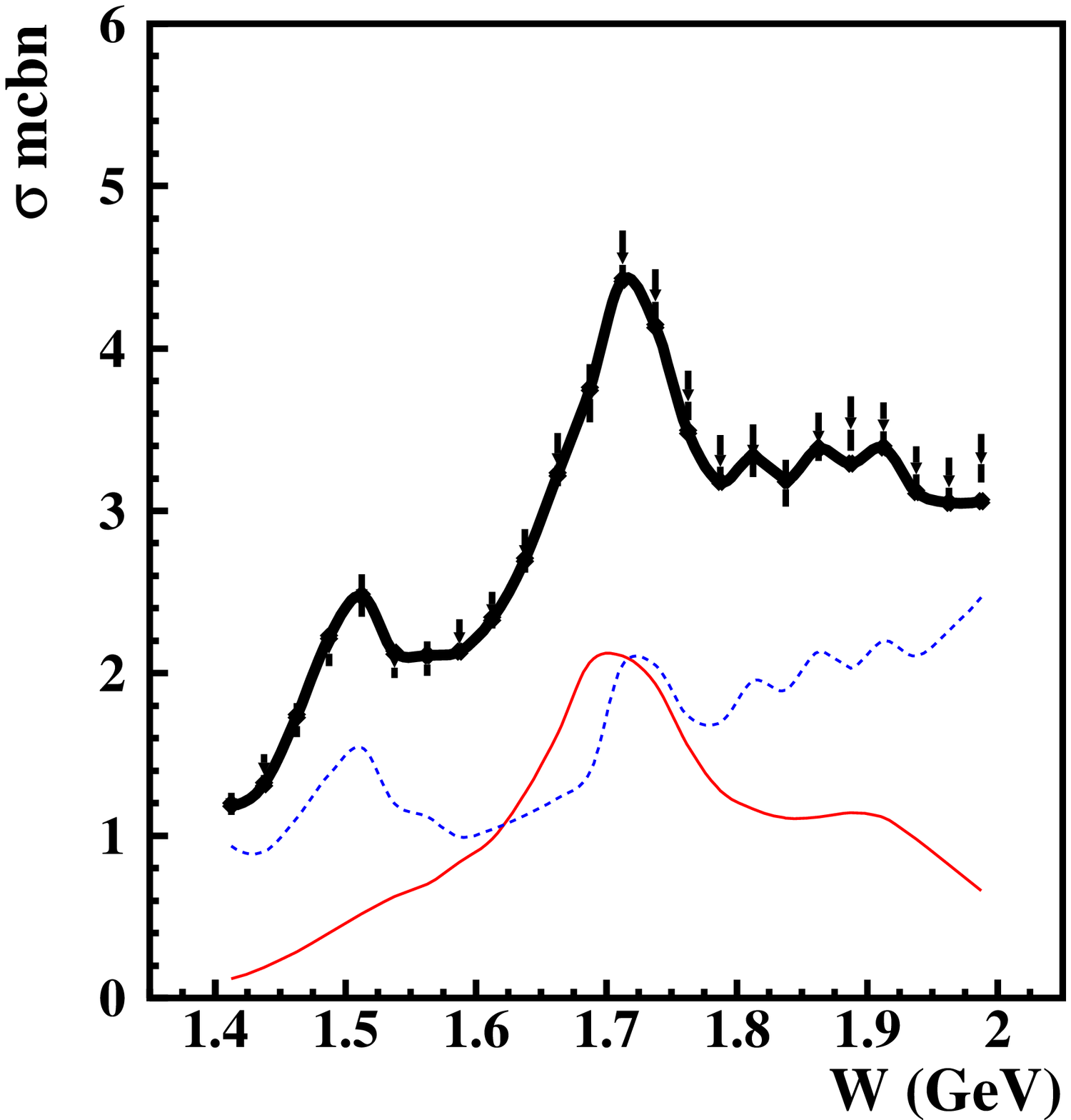}
\includegraphics[width=6cm,clip]{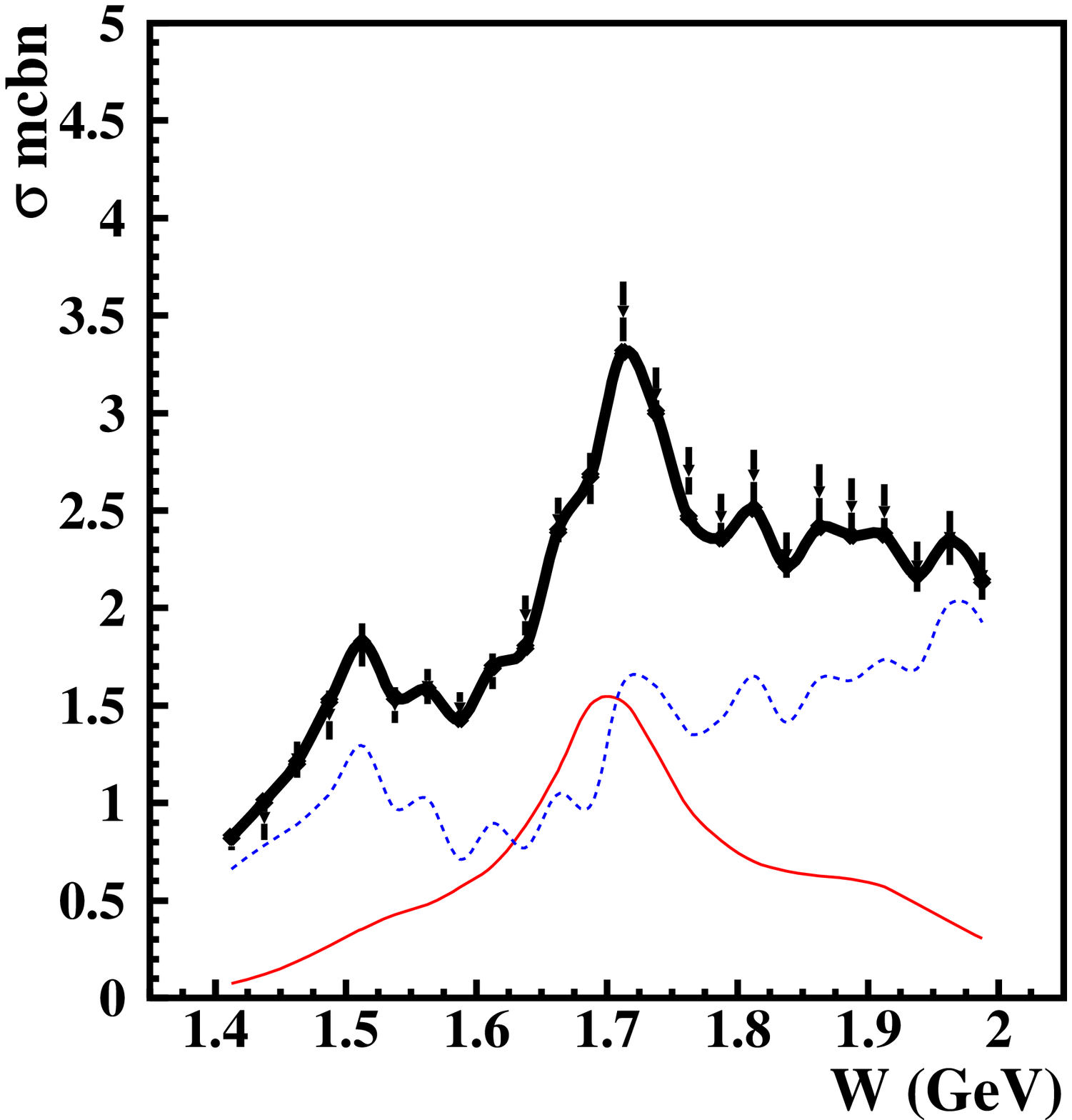} 
\caption{(Color Online) Description of fully integrated $\pi^+\pi^-p$ electroproduction off proton cross sections measured with CLAS \cite{Tr18} 
at $Q^2$ = 2.6 GeV$^2$ (top left),$Q^2$ = 3.2 GeV$^2$ (top right), $Q^2$ = 3.7 GeV$^2$ (bottom). Error bars represent statistical data uncertainties. Computed within the JM model full cross sections, the resonant and non-resonant contributions are shown by black-solid thick, red-solid thin and blue-dotted lines, respectively.   }
\label{tothighq2}
\end{figure}

\begin{figure}[htb!]
\centering
\includegraphics[width=8.1cm,clip]{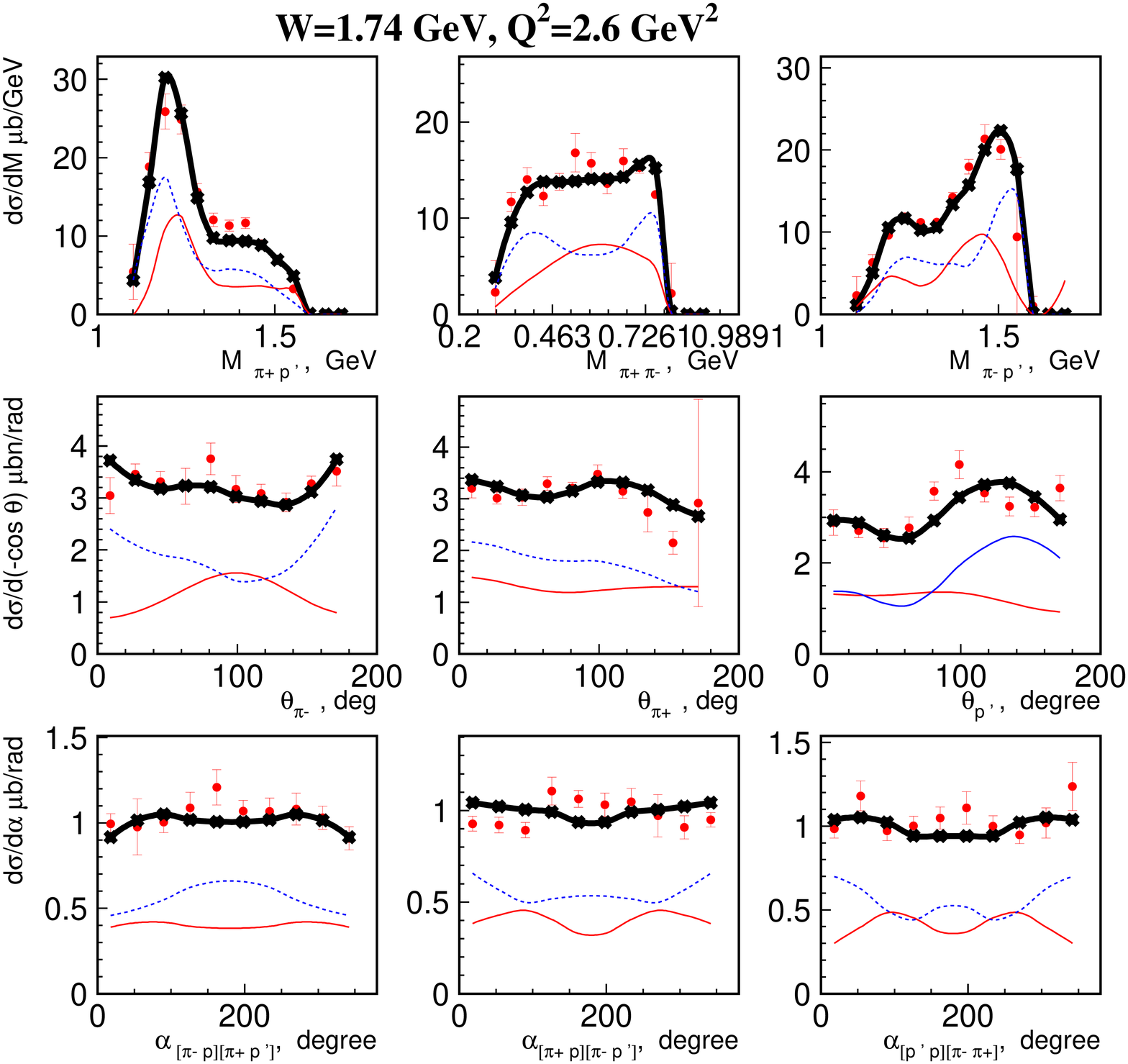}
\includegraphics[width=8.1cm,clip]{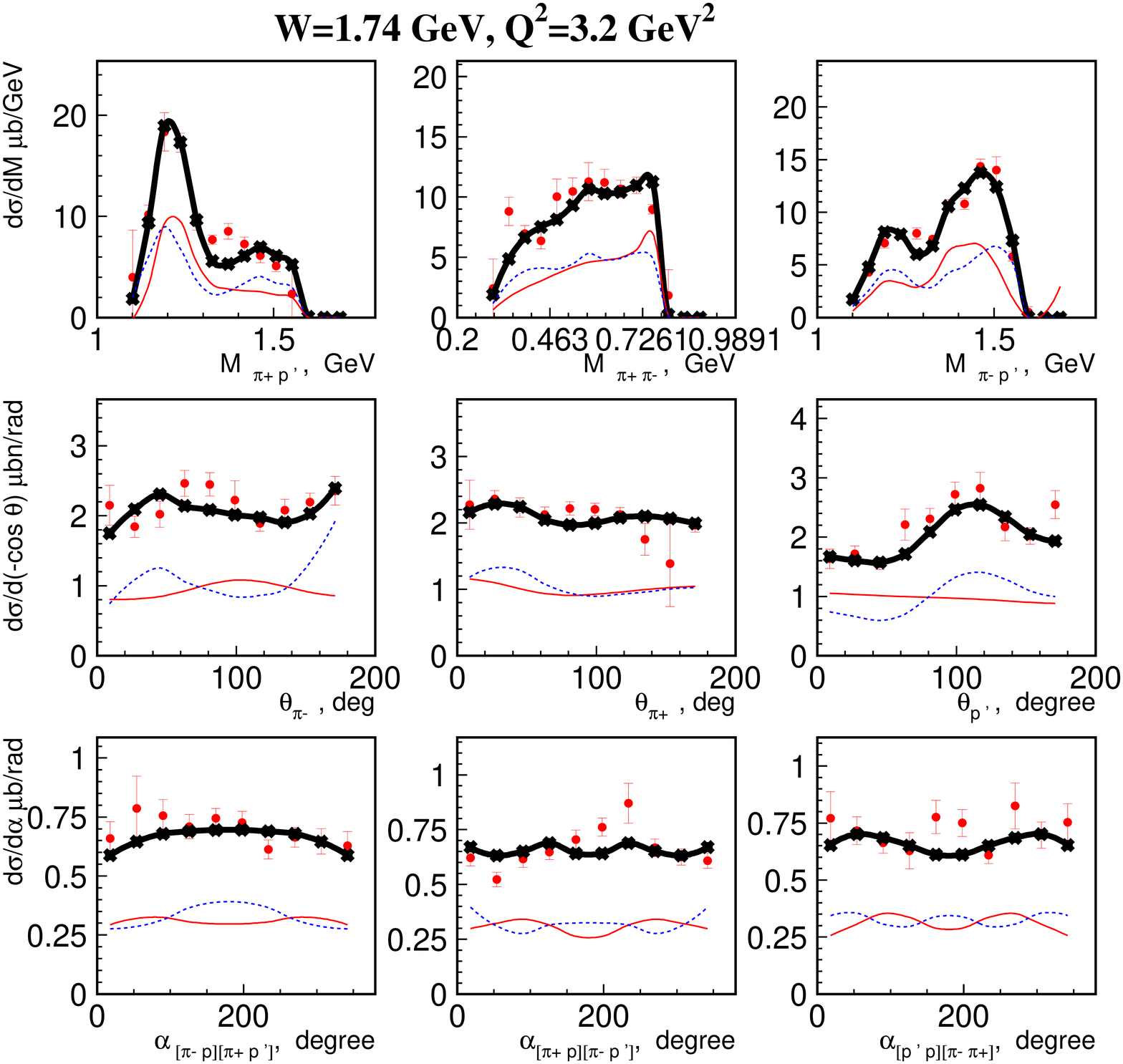}

\caption{(Color Online) Description of nine one-fold differential $\pi^+\pi^-p$ electroproduction off proton cross sections measured with CLAS \cite{Tr18} at $W$=1.74 GeV and $Q^2$ = 2.6 GeV$^2$ (left), $Q^2$ = 3.2 GeV$^2$ (right). Error bars represent statistical data uncertainties. Computed within the JM model full cross sections, the resonant and non-resonant contributions are shown by black-solid thick, red-solid thin and blue-dotted lines, respectively.   }
\label{1diffhighq2}
\end{figure}

Analysis of the CLAS results \cite{Az09}  on the magnetic $p \to \Delta(1232)3/2^+$  form factor within DSE demonstrated for the first time that the masses of dressed quarks are in fact running with quark momentum as predicted by the DSE computations of the dressed quark mass function starting from the QCD-Lagrangian. The DSE evaluation of the magnetic $p \to \Delta(1232)3/2^+$ form factor was carried out by employing the simplified contact qq-interaction (dashed lines in Fig.~\ref{del33p11} (left)) and with the most advanced realistic $qq$-interaction~\cite{Bi17,Ch18} (solid lines in Fig.~\ref{del33p11} (left)). The contact $qq$-interaction produces a dynamically generated dressed quark mass of 
$\approx$400~MeV, that is momentum independent. DSE computations with a realistic $qq$-interaction~\cite{Bi17,Ch18} 
predict a momentum dependent quark mass. The DSE results with a frozen quark mass overestimate the CLAS data at $Q^2 > 1.5$~GeV$^2$. The discrepancies are increasing with $Q^2$. Instead, by employing a quark mass that is running with momentum, the DSE computations offer 
a good description of the experimental results on the $p \to \Delta(1232)3/2^+$ magnetic form factor in the entire range of photon virtualities 0.8 GeV$^2$ $<$ $Q^2$ $<$ 7.0 GeV$^2$. 

Remarkably, a good description of the experimental results on the $p \to \Delta(1232)3/2^+$ transition form factors and the $N(1440)1/2^+$ resonance electrocouplings is achieved with a momentum dependence of the dressed quark mass that is 
{\it exactly the same} as the one employed in the previous evaluations of the elastic electromagnetic nucleon form factors~\cite{Seg14}. This success strongly supports: a) the relevance of dynamical dressed quarks with properties predicted 
by the DSE approach~\cite{Ro17,Bi17,Ch18}, as constituents of the quark core for the structure both of the ground and 
excited nucleon states and b) the capability of the DSE approach~\cite{Seg14,Seg15} to map out the dressed quark mass 
function from the experimental results on the $Q^2$-evolution of the nucleon elastic and $p \to N^*$ electromagnetic transition form factors, or rather $\gamma_vpN^*$ electrocouplings.

The $\gamma_vpN^*$ electrocouplings of many excited nucleon states in the mass range up to 1.7~GeV were evaluated within a novel light front quark model (LFQM)~\cite{Az12,Az17}. This model accounts for the contributions from both the meson-baryon cloud and the quark core, and incorporates the parameterized momentum- dependent quark mass with parameters adjusted to the data on the $Q^2$-evolution of the nucleon elastic form factors. A successful description of the electrocouplings of most resonances in the mass range up to 1.7~GeV was achieved with the same momentum dependent quark mass used for the successful description of the nucleon elastic form factors. A typical example for the description of the $A_{1/2}$ $N(1440)1/2^+$ electrocouplings is shown in Fig.~\ref{del33p11} (right) by the solid line. A successful description of the $\gamma_vpN^*$ electrocouplings for most excited nucleon states in the mass range up to 1.7~GeV offers support for the running dressed quark mass from a framework conceptually different than the DSE approach.

The analysis of the CLAS results on the $\gamma_vpN^*$ electrocouplings of most excited nucleon states in the mass range 
up to 1.7~GeV has revealed the $N^*$ structure for $Q^2 < 5.0$~GeV$^2$ as a complex interplay between an inner core of three dressed quarks and an external meson-baryon cloud~\cite{Bu12,Az12,Az17,Seg15}. The credible DSE evaluation of the quark core contributions to the electrocouplings of the $N(1440)1/2^+$ state~\cite{Seg15} has allowed us 
to infer the meson-baryon cloud contributions to this resonance as the difference between the experimental data on the resonance electrocouplings and the quark core electroexcitation amplitudes computed from DSE, as shown by the shadowed area in Fig.~\ref{del33p11} (right). The relative contributions of the quark core and the meson-baryon cloud depend strongly on the quantum numbers of the excited nucleon state.

\begin{figure}[htbp]
\centering
\includegraphics[width=8.5cm,clip]{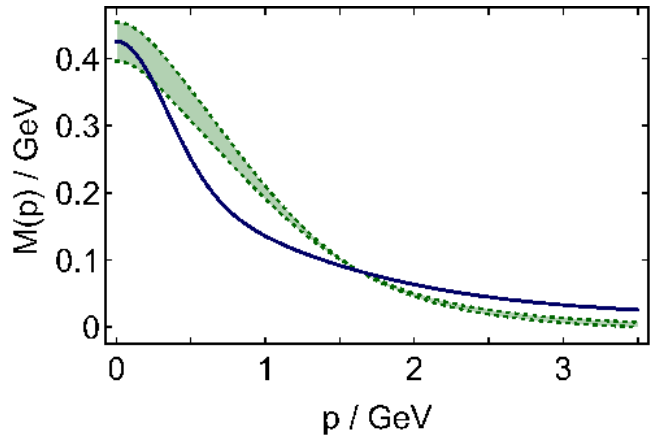}
\caption{Momentum dependence of the dressed quark dynamical mass inferred from the QCD-Lagrangian within the DSE 
approach as the solution of the gap equation tower~\cite{Bi17,Ch18} (green band) and its parameterization (solid line) with parameters fit to the data on the meson and baryon spectrum~\cite{Ch18}.}
\label{qmass} 
\end{figure}

\section{Future prospects in exploration of the nucleon resonance structure from the data on exclusive charged double pion electroproduction off protons}

The kinematics area over W and Q$^2$ covered in the studies of exclusive $\pi^+\pi^- p$ electroproduction off protons is extending gradually with a goal to obtain $\gamma_vpN^*$ electrocouplings of all prominent in this channel excited nucleon states in the full range of photon virtualities covered in the experiments with the CLAS detector of $Q^2$ $<$ 5.0 GeV$^2$. The CLAS data on nine independent one-fold differential and fully integrated charged double pion electroproduction cross sections at 1.4 GeV $<$ W $<$ 2.0 GeV and 2.0 GeV$^2$ $<$ $Q^2$ $<$ 5.0 GeV$^2$ were published in \cite{Is17}. The preliminary results on these observables in the same kinematics area of improved statistical and systematical accuracy have recently become available \cite{Tr18}. The cross sections obtained in \cite{Is17} and \cite{Tr18} by employing different methods for event selection and efficiency evaluation are consistent emphasizing reliability of the both data sets. Phenomenological analysis of these results allows us to extend the kinematic coverage of the JM reaction model for extraction of the nucleon resonance $\gamma_vpN^*$ electrocouplings, their $\pi\Delta$ and $\rho p$ hadronic decay widths, for W up to 2.0 GeV and for $Q^2$ up to 5.0 GeV$^2$. A good description of all measured $\pi^+\pi^-p$ electroproduction off proton cross sections in the aforementioned area of W and $Q^2$ was achieved within the framework of the updated JM model with $\chi^2/d.p.$ in each bin of W and $Q^2$ in the range from 1.2 to 2.9 accounting for the statistical data uncertainties only. Mechanisms already incorporated into the JM model turn out to be sufficient for the data description after modifications of the phenomenological extra-contact terms included into the non-resonant amplitudes of the $\pi\Delta$ meson-baryon channels, phenomenological term for parameterization of $\pi^+ N(1680)5/2^+$ channel amplitude and by implementing fit to the data an additional dependence from the CM-$\pi^-$-emission angle for the non-resonant contributions into $\rho^{0} p$ channel.

Description of fully integrated $\pi^+\pi^-p$ electroproduction off proton cross sections together with the resonant/non-resonant contributions inferred from the data \cite{Is17,Tr18} analysis within the updated JM model is shown in Fig.~\ref{tothighq2}. Representative example for description of nine one-fold differential cross sections in particular bins of W and $Q^2$ along with evaluated resonant and non-resonant contributions are shown in Fig.~\ref{1diffhighq2}.  The resonance contributions remain significant within the entire range of W $<$ 2.0 GeV and $Q^2$ $<$ 5.0 GeV$^2$. The relative resonant contributions in the second and third resonance regions increase with photon virtualities. The shapes of resonant and non-resonant contributions have pronounced differences in the one-fold differential cross sections. Furthermore, the correlations between the shapes of the resonance/non-resonant contributions in different one-fold differential cross sections also demonstrate distinctively different features. All these observations strongly suggest the promising prospects for extraction of $\gamma_vpN^*$ electrocouplings and  the nucleon resonance $\pi\Delta$ and $\rho p$ hadronic decay widths from the $\pi^+\pi^-p$ electroproduction data~\cite{Is17,Tr18}. We are planning to obtain from these data electrocouplings of most excited nucleon state in the mass range up to 2.0 GeV and at photon virtualities from 2.0 GeV$^2$ to 5.0 GeV$^2$. These future results will extend essentially the available information on $\gamma_vpN^*$ electrocouplings summarized in Table~\ref{tab-2}.

After completion of the Jefferson Lab 12~GeV Upgrade Project, the CLAS12 detector operations started successfully in the upgraded 
Hall~B in the Spring of 2018 \cite{clas12}. CLAS12 is the only facility worldwide capable of studying nucleon resonances in the still unexplored ranges of the smallest photon virtualities 0.05~GeV$^2 < Q^2 < 0.5$~GeV$^2$ and the highest photon virtualities up to 12~GeV$^2$ ever achieved in exclusive reaction measurements~\cite{Bu19,Az13}.

In the experiment with CLAS12, we will also search for the hybrid signal as the presence of extra states in the conventional resonance spectrum of $J^P$=1/2$^+$, 3/2$^+$ in the mass range from 2.0~GeV to 2.5~GeV from the data on exclusive 
$KY$ and $\pi^+\pi^-p$ electroproduction off protons~\cite{e12-16-010}. The hybrid nature of the new baryon 
states will be identified by studying for the specific $Q^2$ evolution of their electrocouplings. The studies on $N^*$ structure at low $Q^2$ over the spectrum of all prominent resonances will also continue the long term efforts on the search for the new ``missing" baryon states. The successful description of the combined photo- and electroproduction data in the 
entire range of $Q^2$ by employing a $Q^2$ independent resonance mass, as well as $Q^2$ independent total and partial hadronic decay widths, will validate the presence of new states in the baryon spectrum in a nearly model independent way.

Exploration of the excited nucleon state structure in exclusive $\pi N$, $KY$, and $\pi^+\pi^-p$ electroproduction off protons at 5.0~GeV$^2 < Q^2 < 12.0$~GeV$^2$~\cite{e12-09-003,e12-06-108a,e12-16-010a} is already started in run group A measurements with the CLAS12 detector. For the first time, the electrocouplings of all prominent nucleon resonances will become available at the highest photon virtualities ever achieved in the studies of exclusive electroproduction. These distance scales correspond to the still unexplored regime for $N^*$ electroexcitations where the resonance structure is dominated by the quark core with almost negligible meson-baryon cloud contributions. The foreseen experiments offer direct access to the properties of dressed quarks inside $N^*$ states of different quantum numbers. Consistent results on the dressed quark mass function derived from independent analyses of the data on the $\gamma_vpN^*$ electrocouplings of the resonances with distinctively different structure, such as radial excitations, spin-isospin flip, and orbital excitations, will validate the credible access to this fundamental ingredient of strong QCD from the experimental data. The expected data on the $\gamma_vpN^*$ electrocouplings will provide  access to the dressed quark mass function in the range of quark momenta up to 1.5~GeV, where the transition from the quark-gluon confinement to the pQCD regimes of the strong interaction takes full effect, as is shown in Fig.~\ref{qmass}. Exploring 
the dressed quark mass function at these distances will allow us to address the most challenging open problems 
of the Standard Model on the nature of $>$98\% of hadron mass and quark-gluon confinement~\cite{BurkertRoberts,Az13}.

\bigskip

The U.S. Government retains a non-exclusive, paid-up, irrevocable, world-wide license to publish or reproduce this
manuscript for U.S. Government purposes. This material is based upon work supported by the U.S. Department
of Energy, Office of Science, Office of Nuclear Physics under contract DE-AC05-06OR23177. 




\end{document}